\documentclass[%
reprint,
superscriptaddress,
amsmath,amssymb,
aps,
]{revtex4-2}

\usepackage{blindtext}
\usepackage{graphicx}% Include figure files
\usepackage{dcolumn}% Align table columns on decimal point
\usepackage{bm}% bold math
\usepackage[colorlinks=true,
            linkcolor=black,
            citecolor=black,
            urlcolor=black]{hyperref}% add hypertext capabilities
\usepackage[mathlines]{lineno}% Enable numbering of text and display math with alternating line numbers
\usepackage{lipsum}
\usepackage{booktabs}
\usepackage{layouts}
\usepackage{multirow}
\usepackage{xcolor}  % Add this in the preamble
\usepackage{soul}
\usepackage{orcidlink}

\makeatletter
\renewcommand{\fnum@figure}{FIG.~\thefigure}
\renewcommand{\fnum@table}{TABLE~\thetable}

\makeatother

\def\SM{Supplemental Material\,}
\def\MK{M_{K^\text{-}}}

\begin{document}

\preprint{APS/123-QED}

\title{
Precision Test of Bound-State QED at Intermediate-Z with Kaonic Neon
}
%Kaonic Neon X-Ray Transitions as Probes for Bound-State QED Tests

% ----- Corresponding author -----
\author{Manti S.}
\thanks{Mail: \texttt{Simone.Manti@lnf.infn.it} (Corresponding Author)}
\affiliation{Laboratori Nazionali di Frascati INFN, Frascati, Italy}

% ----- Remaining authors (in original order) -----
\author{Sgaramella F.}
\affiliation{Laboratori Nazionali di Frascati INFN, Frascati, Italy}

\author{Abbene L.}
\affiliation{Department of Physics and Chemistry (DiFC), Emilio Segrè, University of Palermo, Palermo, Italy}
\affiliation{Laboratori Nazionali di Frascati INFN, Frascati, Italy}

\author{Amsler C.}
\affiliation{Stefan Meyer Institute for Subatomic Physics, Vienna, Austria}

\author{Artibani F.}
\affiliation{Laboratori Nazionali di Frascati INFN, Frascati, Italy}
\affiliation{Università degli studi di Roma Tre, Dipartimento di Fisica, Roma, Italy}

\author{Bazzi M.}
\affiliation{Laboratori Nazionali di Frascati INFN, Frascati, Italy}

\author{Borghi G.}
\affiliation{Politecnico di Milano, Dipartimento di Elettronica, Informazione e Bioingegneria, Milano, Italy}
\affiliation{INFN Sezione di Milano, Milano, Italy}

\author{Bosnar D.}
\affiliation{Department of Physics, Faculty of Science, University of Zagreb, Zagreb, Croatia}
\affiliation{Laboratori Nazionali di Frascati INFN, Frascati, Italy}

\author{Bragadireanu M.}
\affiliation{Horia Hulubei National Institute of Physics and Nuclear Engineering (IFIN-HH), Măgurele, Romania}

\author{Buttacavoli A.}
\affiliation{Department of Physics and Chemistry (DiFC), Emilio Segrè, University of Palermo, Palermo, Italy}
\affiliation{Laboratori Nazionali di Frascati INFN, Frascati, Italy}

%\author{Cargnelli M.}
%\affiliation{Stefan-Meyer-Institut für Subatomare Physik, Vienna, Austria}

\author{Carminati M.}
\affiliation{Politecnico di Milano, Dipartimento di Elettronica, Informazione e Bioingegneria, Milano, Italy}
\affiliation{INFN Sezione di Milano, Milano, Italy}

\author{Clozza A.}
\affiliation{Laboratori Nazionali di Frascati INFN, Frascati, Italy}

\author{Clozza F.}
\affiliation{Laboratori Nazionali di Frascati INFN, Frascati, Italy}
\affiliation{Università degli studi di Roma Tor Vergata, Dipartimento di Fisica, Roma, Italy}

\author{Del Grande R.}
\affiliation{Faculty of Nuclear Sciences and Physical Engineering, Czech Technical University in Prague, Břehovà 7, 115 19, Prague, Czech Republic}
\affiliation{Laboratori Nazionali di Frascati INFN, Frascati, Italy}

\author{De Paolis L.}
\affiliation{Laboratori Nazionali di Frascati INFN, Frascati, Italy}

\author{Dulski K.}
\affiliation{Laboratori Nazionali di Frascati INFN, Frascati, Italy}
\affiliation{Faculty of Physics, Astronomy, and Applied Computer Science, Jagiellonian University, Kraków, Poland}
\affiliation{Center for Theranostics, Jagiellonian University, Krakow, Poland}

\author{Fabbietti L.}
\affiliation{Physik Department E62, Technische Universität München, James-Franck-Straße 1, 85748 Garching, Germany}

\author{Fiorini C.}
\affiliation{Politecnico di Milano, Dipartimento di Elettronica, Informazione e Bioingegneria, Milano, Italy}
\affiliation{INFN Sezione di Milano, Milano, Italy}

\author{Friščić I.}
\affiliation{Department of Physics, Faculty of Science, University of Zagreb, Zagreb, Croatia}

\author{Guaraldo C.}
\thanks{Deceased}
\affiliation{Laboratori Nazionali di Frascati INFN, Frascati, Italy}

\author{Iliescu M.}
\affiliation{Laboratori Nazionali di Frascati INFN, Frascati, Italy}

\author{Indelicato P.}
\affiliation{Laboratoire Kastler Brossel, Sorbonne Université, CNRS, ENS-PSL Research University, Collège de France, Case 74; 4, place Jussieu, F-75005 Paris, France}

\author{Iwasaki M.}
\affiliation{RIKEN, Tokyo, Japan}

\author{Khreptak A.}
\affiliation{Faculty of Physics, Astronomy, and Applied Computer Science, Jagiellonian University, Kraków, Poland}
\affiliation{Center for Theranostics, Jagiellonian University, Krakow, Poland}
\affiliation{Laboratori Nazionali di Frascati INFN, Frascati, Italy}

\author{Marton J.}
\affiliation{Stefan Meyer Institute for Subatomic Physics, Vienna, Austria}
\affiliation{Atominstitut, Technische Universität Wien, Stadionallee 2, 1020 Vienna, Austria}

\author{Moskal P.}
\affiliation{Faculty of Physics, Astronomy, and Applied Computer Science, Jagiellonian University, Kraków, Poland}
\affiliation{Center for Theranostics, Jagiellonian University, Krakow, Poland}

\author{Napolitano F.}
\affiliation{Via A. Pascoli 06123, Perugia (PG), Italy, Dipartimento di Fisica e Geologia, Università degli studi di Perugia}
\affiliation{INFN Sezione di Perugia, Via A. Pascoli, 06123 Perugia, Italia}

\author{Niedźwiecki S.}
\affiliation{Faculty of Physics, Astronomy, and Applied Computer Science, Jagiellonian University, Kraków, Poland}
\affiliation{Center for Theranostics, Jagiellonian University, Krakow, Poland}

\author{Ohnishi H.}
\affiliation{Research Center for Accelerator and Radioisotope Science (RARiS), Tohoku University, Sendai, Japan}

\author{Piscicchia K.}
\affiliation{Centro Ricerche Enrico Fermi, Museo Storico della Fisica e Centro Studi e Ricerche ``Enrico Fermi", Roma, Italy}
\affiliation{Laboratori Nazionali di Frascati INFN, Frascati, Italy}

\author{Principato F.}
\affiliation{Department of Physics and Chemistry (DiFC), Emilio Segrè, University of Palermo, Palermo, Italy}
\affiliation{Laboratori Nazionali di Frascati INFN, Frascati, Italy}

\author{Scordo A.}
\affiliation{Laboratori Nazionali di Frascati INFN, Frascati, Italy}

\author{Silarski M.}
\affiliation{Faculty of Physics, Astronomy, and Applied Computer Science, Jagiellonian University, Kraków, Poland}

\author{Sirghi D.}
\affiliation{Centro Ricerche Enrico Fermi, Museo Storico della Fisica e Centro Studi e Ricerche ``Enrico Fermi", Roma, Italy}
\affiliation{Laboratori Nazionali di Frascati INFN, Frascati, Italy}
\affiliation{Horia Hulubei National Institute of Physics and Nuclear Engineering (IFIN-HH), Măgurele, Romania}

\author{Sirghi F.}
\affiliation{Laboratori Nazionali di Frascati INFN, Frascati, Italy}
\affiliation{Horia Hulubei National Institute of Physics and Nuclear Engineering (IFIN-HH), Măgurele, Romania}

\author{Skurzok M.}
\affiliation{Faculty of Physics, Astronomy, and Applied Computer Science, Jagiellonian University, Kraków, Poland}
\affiliation{Center for Theranostics, Jagiellonian University, Krakow, Poland}
\affiliation{Laboratori Nazionali di Frascati INFN, Frascati, Italy}

\author{Spallone A.}
\affiliation{Laboratori Nazionali di Frascati INFN, Frascati, Italy}

\author{Sommerfeldt J.}
\affiliation{Laboratoire Kastler Brossel, Sorbonne Université, CNRS, ENS-PSL Research University, Collège de France, Case 74; 4, place Jussieu, F-75005 Paris, France}

\author{Toscano L.G.}
\affiliation{Politecnico di Milano, Dipartimento di Elettronica, Informazione e Bioingegneria, Milano, Italy}
\affiliation{INFN Sezione di Milano, Milano, Italy}

%\author{Tüchler M.}
%\affiliation{Stefan-Meyer-Institut für Subatomare Physik, Vienna, Austria}

\author{Toho K.}
\affiliation{Research Center for Accelerator and Radioisotope Science (RARiS), Tohoku University, Sendai, Japan}
\affiliation{Laboratori Nazionali di Frascati INFN, Frascati, Italy}

\author{Vazquez Doce O.}
\affiliation{Laboratori Nazionali di Frascati INFN, Frascati, Italy}

% \author{Widmann E.}
% \affiliation{Stefan-Meyer-Institut für Subatomare Physik, Vienna, Austria}

\author{Zmeskal J.}
\thanks{Deceased}
\affiliation{Stefan Meyer Institute for Subatomic Physics, Vienna, Austria}

\author{Curceanu C.}
\affiliation{Laboratori Nazionali di Frascati INFN, Frascati, Italy}

\collaboration{SIDDHARTA-2 Collaboration}%\noaffiliation

\date{\today}% It is always \today, today,

\begin{abstract}
We report Dirac-Fock calculations of transition energies for kaonic neon (KNe). For the most intense line, the 7–6 transition, the calculated energy is $9450.28$ eV, which includes a bound-state QED (BSQED) contribution of $12.66$ eV. This is in excellent agreement with the recent SIDDHARTHA-2 measurement at DA$\Phi$NE of $9450.23 \pm 0.37\,(\mathrm{stat.}) \pm 1.50\,(\mathrm{syst.})$ eV. With the QED shift far exceeding experimental uncertainty, these results establish kaonic atoms as powerful platforms for precision tests of BSQED in intermediate-Z systems.
\end{abstract}

% \keywords{Suggested keywords}%Use showkeys class option if keywordExperimental Setup and Methods%                             %display desired
\maketitle
%
%\tableofcontents
% \printinunitsof{in}\prntlen{\textwidth}  
%
%%%%%%%%%%%%%%%%%%%%%%%%%%%%%%%%%%%%%%%%%%%%
\section{Introduction}
%%%%%%%%%%%%%%%%%%%%%%%%%%%%%%%%%%%%%%%%%%%%
%
% QED
Quantum Electrodynamics (QED) is the most accurate theory describing the interaction between charged particles and photons within the framework of quantum field theory.
% BSQED
For instance, in the context of Bound-State QED (BSQED) \cite{indelicato_electron_1999,indelicato_introduction_2016}, which utilizes QED to describe interactions in bound atomic systems, theoretical predictions achieve parts-per-billion (ppb) accuracy for the 2S–1S transition in hydrogen \cite{parthey_improved_2011}. However, for intermediate atomic numbers (Z), the theoretical calculations face limitations in reaching similar precision, as the perturbative expansion scales with $(\alpha Z)$, with $\alpha$ the fine-structure constant, leading to convergence issues for high-Z elements.
% HCIs
To explore the intermediate-$Z$ case, tests of BSQED are typically conducted using Highly Charged Ions (HCIs) \cite{indelicato_qed_2019,shabaev_stringent_2018,morgnerStringentTestQED2023} confined in cyclotrons and ion traps, enabling the exploration of QED effects in strong electric fields  \cite{ullmann_high_2017}. Nevertheless, in such systems, the precision of QED contributions is comparable in magnitude to Finite Nuclear Size (FNS) effects, thereby limiting the overall experimental accuracy \cite{volotka_progress_2013}.
% Exotic Atoms
Recently, a paradigm shift has occurred in BSQED tests at intermediate-$Z$, favoring the use of exotic atoms \cite{paul_testing_2021}. These systems, where a negatively charged exotic particle (e.g. $\mu^{-}$, $\bar{p}$, $K^{-}$, $\pi^{-}$) replaces an electron, enable probing regions closer to the nucleus due to the larger mass of the exotic particle compared to the electron, where BSQED effects are highly enhanced \cite{gotta_precision_2004}.
% Cascade
Historically, studies of exotic atoms were constrained to solid targets \cite{jeckelmann_new_1986}, where electron refilling effects often left residual electrons bound to the atom, complicating the interpretation of spectroscopic data \cite{simons_exotic_1994}. Recent advancements in detection and accelerator technologies have enabled investigations using low-pressure gas targets, significantly reducing electron refilling and allowing for cleaner observations of atomic transitions \cite{sgaramella_high_2025,okumura_proof--principle_2023,okumura_few-electron_2025}.
% Kaonic Atoms
To date, BSQED tests in exotic atoms have predominantly utilized muonic, pionic and antiprotonic atoms \cite{paul_testing_2021}. In contrast, negatively charged kaons present unique advantages for such studies \cite{curceanu_modern_2019}. Being spin-0 particles, kaons are not subject to hyperfine splitting, unlike muons, simplifying the spectral analysis. Additionally, having an intermediate mass between that of pions, muons, and antiprotons, kaons could provide for BSQED studies a more complete picture across the entire mass range. Moreover, specific transitions within the kaonic atom cascade can be selected to minimize the influence of FNS effects and residual electrons.
A notable challenge in employing kaons for precision measurements is the ``kaon mass problem''. Several discrepancies exist among measurements of the kaon mass \cite{backenstoss_k_1973,barkov_charged_1979,cheng_k_1975,lum_kaonic_1981}, with the primary disagreement occurring between the two most precise results \cite{denisov_new_1991,gall_precision_1988}. This has led to a persistent uncertainty of approximately 30 parts per million (ppm), as reported by the Particle Data Group (PDG) \cite{navas_review_2024}. Resolving this issue is crucial for enhancing the accuracy of BSQED tests involving kaonic atoms.
% MCDFGME
The majority of these studies were conducted prior to 1990, when accurate methods, particularly for addressing electron screening effects, were not yet well established.
%and various sources of uncertainties in determining the final mass had not been clearly separated. 
Nowadays, state-of-the-art Multiconfiguration Dirac-Fock (MCDF) methods \cite{mallow_dirac-fock_1978} offer the possibility to take into account all the different effects, potentially uniquely linking and determining the kaon mass from the experiment.
% SIDDHARTA-2
Recently, the SIDDHARTA-2 collaboration at the DA$\Phi$NE collider of the National Laboratories of Frascati (INFN-LNF) in Italy, in preparation for the subsequent measurement on kaonic deuterium, successfully performed high-precision X-ray spectroscopy measurements of kaonic neon (KNe with K for K$^-$) \cite{sgaramella_high_2025}. Such gaseous systems exhibit several transitions with high yields ($>\!\!30\%$) in the X-ray range corresponding to high-n levels. Thanks to the performed optimization and the use of Silicon Drift Detectors (SDDs) \cite{miliucci_silicon_2021}, sub-eV precision was achieved on the measurement of those lines.
% In this article...
In this Letter, we compare the recent SIDDHARTA‑2 KNe measurements \cite{sgaramella_high_2025} with state‑of‑the‑art MCDF calculations. We demonstrate excellent agreement between experiment and theory, underscoring the sensitivity of KNe transitions, and kaonic atoms more broadly, to BSQED effects. We also validate the result against the electron screening effect and kaon‑mass dependence, and finally show that KNe offers a viable route to refining the charged kaon mass with high precision.
%
%%%%%%%%%%%%%%%%%%%%%%%%%%%%%%%%%%%%%%%%%%%%
\section{Methods}
%%%%%%%%%%%%%%%%%%%%%%%%%%%%%%%%%%%%%%%%%%%%
%
% Experimental Details
KNe data were recorded during the 2023 SIDDHARTA-2 run at the DA$\Phi$NE \cite{milardi_preparation_2018,milardi_dane_2021,milardi_et_dafne_2024} \mbox{accelerator} at INFN-LNF. The sample used in this Letter corresponds to an integrated luminosity of $150\;\mathrm{pb}^{-1}$, representing a 20\% increase in statistics compared to our previous result \cite{sgaramella_high_2025}. %and thus allowing for improved statistical precision. 
Further experimental details on the SIDDHARTA-2 setup and KNe measurement are provided in Refs. \cite{sirghi_siddharta-2_2024} and \cite{sgaramella_high_2025}, respectively. 
% Fit Details: Likelihood fit, Gaussian peaks, Fano and Noise, Background
To extract the transition energies from the X-ray spectrum, we performed a maximum likelihood fit in which the observed spectral peaks were modeled using gaussian functions. Each gaussian function was explicitly parameterized by the SDDs resolution, incorporating both the Fano factor and electronic noise, consistent with the methodology described in our previous study \cite{miliucci_silicon_2021}. %Gaussian components were iteratively added to the fit model until the points in the pull plot consistently lay within the $\pm3$ range, ensuring the goodness of fit.
% Isotopic
To account for the presence of different neon isotopes in our spectral analysis, we modeled the observed lines using double gaussian functions. The amplitudes of the two components were constrained based on the natural isotopic abundances of neon \cite{audi_ame2003_2003}, namely a \textsuperscript{22}Ne / \textsuperscript{20}Ne ratio of 9.78\%. The inclusion of \textsuperscript{21}Ne, which has a natural abundance of 0.27\%, as well as the uncertainty in the isotopic abundance ratios, does not affect the fit results within the sensitivity of this study.
% Systematics
Systematic uncertainties on the experimental transition energies arise from both the energy stability over time and the calibration procedure, as previously described \cite{sgaramella_measurements_2023}.
% MCDF calculations
To compute the X-ray transition energies, we utilized the MCDFGME code (version 2025.1) developed by Desclaux and Indelicato \cite{mallow_dirac-fock_1978,santos_x-ray_2005,indelicato_qed_2019}, with the 2018 CODATA fundamental constants \cite{tiesinga_codata_2021}. For each transition, the code solves the Klein-Gordon equation for both initial and final states of the kaonic atom, and QED contributions are evaluated with vacuum polarization at all-order \cite{schmidt_higher-order_1989,soff_vacuum_1988}.
%, dominated by vacuum polarization for kaonic atoms, are incorporated non-self-consistently to first and second order and included in the final transition energies. 
In all calculations, we consider only transitions between high-$n$ circular states ($\ell=n-1$). FNS effects are neglected by adopting the point-nucleus approximation, as their impact on transition energies is limited to a few meV, as reported in the Supplemental Material \cite{supp}. Electron screening effects are evaluated by including a single electron in the $1s$ orbital of Ne, constructing the total wave function from the electron and kaon. Unless otherwise specified, we adopt the reference kaon mass from the PDG \cite{navas_review_2024} of $\MK=493.677$ MeV.
% Isotopic Shift
The MCDFGME calculations were used further to constrain the fit of the neon isotopic peaks: the centroid of the K\textsuperscript{22}Ne peak was fixed to the isotopic shift predicted by the calculation, similar to $\mu$Ne \cite{okumura_proof--principle_2023}, and the resolution of the peak was scaled accordingly. In the \SM \cite{supp}, we validated this approach by assessing the sensitivity of the isotopic shifts for our lines to variations in the kaon mass and found that changing the kaon mass by $\pm$100 keV (more than seven times the current PDG uncertainty), leads to a shift of less than 10 meV (see Fig. S1) in the isotopic shift values. This confirms the robustness of our spectral modeling to fit different isotopic contributions.
% Radiative and Auger Rates
Radiative and Auger transition rates were calculated using the hydrogenic wavefunctions and standard analytical formulae widely used in exotic atom studies \cite{burbidge_mesonic_1953}. While the development of a dedicated MCDFGME treatment for Auger transitions is planned for future work, we validate the hydrogenic approach by comparing dipole radiative rates from the hydrogenic approximation and MCDFGME calculations. As detailed in the \SM \cite{supp}, where we find that the relative error for high-$n$ transitions ($n>5$) is below 1\% (see Fig. S2). %Explicit formulas and further details are also provided in the SM.
% Kaon Mass
Additionally, to further highlight the potential of KNe measurement, we extract the kaon mass directly from the measured X-ray transitions. This is accomplished by iteratively adjusting the kaon mass in the theoretical calculation until agreement with the experimental value is achieved, following the method of Gall \cite{gall_precision_1988}.
%, in formulae as:
%
% \begin{equation}
% \mathrm{M}_{K^-}^{\prime} = M_{K^-} \frac{E_{if}^{exp.}}{E_{if}^{calc.}} \qquad
% \delta M_{K^-} = M_{K^-} \frac{\delta E_{if}^{}}{E_{if}^{}}.
% \label{eq:gal}
% \end{equation}
%
% with the uncertainty over the kaon mass being:
% %
% \begin{equation}
% \delta M_{K^-} = M_{K^-} \frac{\delta E_{if}^{}}{E_{if}^{}}
% \label{eq:smk}
% \end{equation}
%
Here, we terminated the iterative procedure once the kaon mass change between successive steps fell below 0.5 keV. Convergence curves related to this approach for the KNe transitions of this work are provided in the \SM \cite{supp} (see Fig. S4).
%
%%%%%%%%%%%%%%%%%%%%%%%%%%%%%%%%%%%%%%%%%%%%
\section{Results and Discussion}
%%%%%%%%%%%%%%%%%%%%%%%%%%%%%%%%%%%%%%%%%%%%
We now report the experimental transition energies of KNe measured by SIDDHARTA-2 and compare them with the MCDFGME calculations. We then discuss the significance of these results for BSQED studies and their implications for a precise determination of the kaon mass.
\subsection{KNe Spectrum}
\begin{figure*}[t!]
    \centering
    \includegraphics[width=0.9\linewidth]{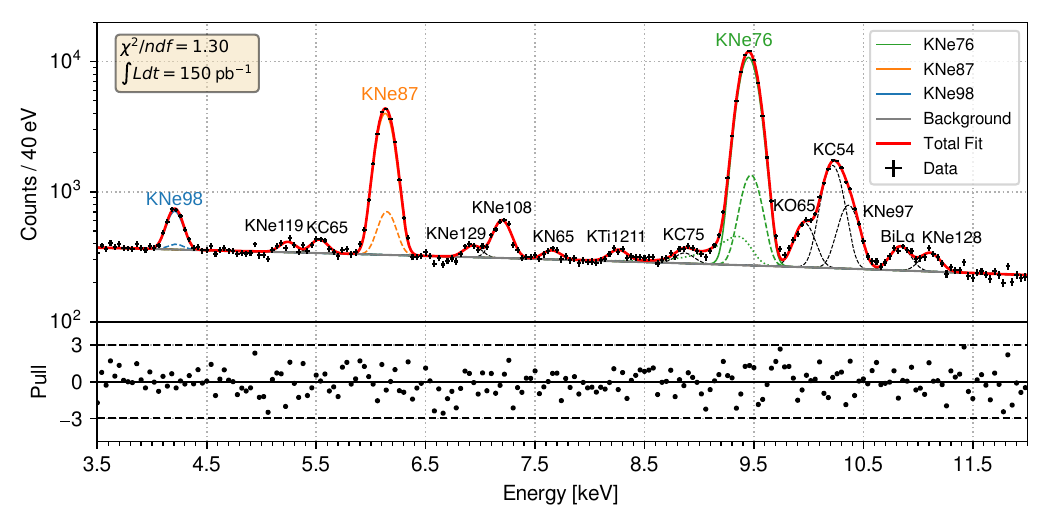}
    \caption{Fit of the KNe X-ray spectrum (top panel) in the 3.5-12 keV range, showing the counts (black), the total fit (red), individual peak contributions (dashed black), and the background (gray). The main lines of KNe are highlighted: KNe76 (green), KNe87 (orange), and KNe98 (blue), each exhibiting a double component due to the presence of two different neon isotopes (\textsuperscript{20}Ne and\textsuperscript{22}Ne). For the KNe76 peak, the tail contribution is also shown. The pull plot (bottom panel) displays the fit residuals normalized by the count errors.}
    \label{fig:spectrum}
\end{figure*}
\begin{figure}[t!]
    \centering
    \includegraphics[width=0.9\columnwidth]{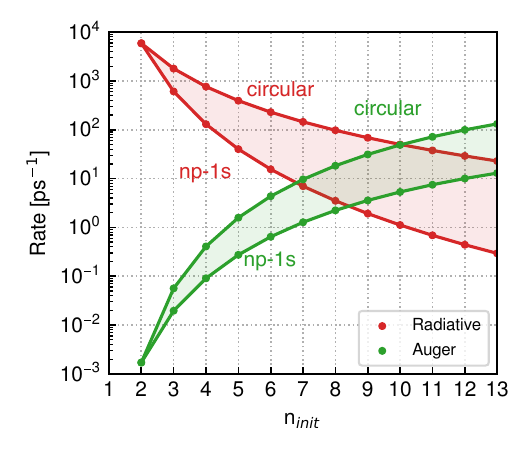}
    \caption{Radiative and Auger transition rates for KNe as a function of the initial principal quantum number $n$. Rates for circular ($\ell = n-1$) and $np \rightarrow 1s$ transitions are plotted as lines, while intermediate transitions fall in the corresponding shaded regions.}
    \label{fig:rates}
\end{figure}

%
% Spectrum Describe peaks
The experimental spectrum of KNe, shown in Fig. \ref{fig:spectrum}, features three dominant peaks corresponding to the 7–6, 8–7, and 9–8 transitions, observed at approximately 9.4 keV, 6.1 keV, and 4.2 keV, respectively. In addition, several lower-intensity peaks are present which include transitions with $\Delta n>1$, such as 10-8, 11–9, 12–9, and 12–8. The spectrum also reveals X-ray lines from other kaonic atoms formed in the setup environment, including carbon, oxygen, nitrogen, and titanium. A distinct peak at 10.8 keV is attributed to the bismuth L$\alpha$ transition, originating from bismuth in the SDD ceramic materials \cite{miliucci_silicon_2021}. The identification and relative intensities of these contaminant lines are consistent with our previous observations \cite{sgaramella_measurements_2023}, and their contributions have been carefully included in the analysis.
\begin{table*}[t!]
    \centering
    \caption{Experimental transition energies $E_{if}^{\text{(exp.)}}$ for KNe obtained from the fit, including their statistical $\delta E_{if}^{\text{(stat.)}}$ and systematic $\delta E_{if}^{\text{(sys.)}}$ uncertainties, along with the calculated values $E_{if}^{\text{(calc.)}}$. The table also shows the QED contributions $E_{if}^{\text{(QED)}}$, isotopic $\Delta E_{if}^\text{(isot.)}$ and electron screening energy shifts $\Delta E_{if}^{(\text{screen.})}$ and uncertainty due to the PDG kaon mass $\Delta E_{if}^\text{(PDG)}$. All energies are given in eV.}
    \vskip10pt
    \begin{ruledtabular}
    \begin{tabular}{ccccccccccc}
        % \toprule
        \textbf{\boldmath Transition} & 
        \textbf{\boldmath $E_{if}^{\text{(exp.)}}$} & 
        \textbf{\boldmath $\delta E_{if}^{\text{(stat.)}}$} &
        \textbf{\boldmath $\delta E_{if}^{\text{(sys.)}}$} &
        \textbf{\boldmath $E_{if}^{\text{(calc.)}}$} & 
        \textbf{\boldmath $E_{if}^{\text{(QED)}}$} & 
        \textbf{\boldmath $E_{if}^{\text{(QED1)}}$} & 
        \textbf{\boldmath $E_{if}^{\text{(QED2)}}$} &
        \textbf{\boldmath $\Delta E_{if}^\text{(isot.)}$} &
        \textbf{\boldmath $\Delta E_{if}^{(\text{screen.})}$} &
        \textbf{\boldmath $\Delta E_{if}^\text{(PDG)}$}\\
        \midrule
        9l-8k  & 4206.97 & 3.43 & 2.00 & 4201.45 &  2.09 &  2.07 & 0.02 & 9.90 & -0.38 & 0.11 \\
        8k-7i  & 6130.57 & 0.65 & 1.50 & 6130.31 & 5.09 &  5.05 & 0.04 & 14.45 &  -0.27 & 0.16 \\
        7i-6h  & 9450.23 & 0.37 & 1.50 & 9450.28 & 12.66 & 12.56 & 0.10 & 22.28 &  -0.18 & 0.24 \\
        6h-5g\footnote{Ref. \cite{sgaramella_high_2025}}  & 15673.30 & 0.52 & 9.00 & 15685.39 & 32.75 & 32.51 & 0.24 & 37.01 & -0.11 & 0.40 \\
        % \bottomrule
    \end{tabular}
    \end{ruledtabular}
    \label{tab:results}
\end{table*}
%
% Rates Radiative and Auger
The observed KNe transitions are consistent with expectations based on the competition between radiative and Auger decay channels during the cascade, as shown in Fig. \ref{fig:rates}. A qualitative analysis of the calculated rates predicts at which levels of the cascade radiative transitions begin to dominate over Auger processes. Specifically, for KNe, radiative decay becomes the dominant channel starting from an initial principal quantum number of approximately $n_\text{init}$=$9$ of the cascade (see Fig. \ref{fig:rates}). Transitions originating from levels below this threshold begin to appear in the spectrum, marking the progressive dominance with increasing yield of radiative emission over Auger de-excitation. %A comprehensive quantitative analysis of the measured yields would require detailed cascade simulations that are currently under development and will be addressed in forthcoming work.
%
% Results of Experimental fit
A fit was performed to extract the transition energies of the observed peaks (see Fig. \ref{fig:spectrum}). The fitting procedure was refined relative to our previous study \cite{sgaramella_high_2025} to enable a direct comparison with theory, specifically by decoupling the contributions from the two neon isotopes (\textsuperscript{20}Ne and \textsuperscript{22}Ne). Due to the high yield a tail function to the 7-6 line was added, with tail parameters fixed from our previous work \cite{sgaramella_measurements_2023}. The fit was restricted to the 3.5–12 keV range, as convergence becomes challenging above this region, particularly for the 6–5 line at 15.7 keV, which closely overlaps with 8–6 line, like in kaonic nitrogen \cite{ishiwatari_kaonic_2004}, unless the relative yields are constrained. 
%The relative yield of 6–5 and 8–6 could, in principle, be constrained using detailed cascade models \cite{koike_electron_2005}; this is left for future work.
A linear background model was adopted, since more complex forms did not yield a significant improvement in the reduced chi-squared value. Isotopic peaks corresponding to the same transition were constrained to improve fit stability, as detailed in the Methods section. 
Results for the principal KNe transitions are reported in Table \ref{tab:results}, with their statistical and systematic uncertainties. The value for 6–5 is also included in the table for completeness and as a reference for the calculations.
\subsection{BSQED}
\begin{figure}[t!]
    \centering
    \includegraphics[width=0.99\columnwidth]{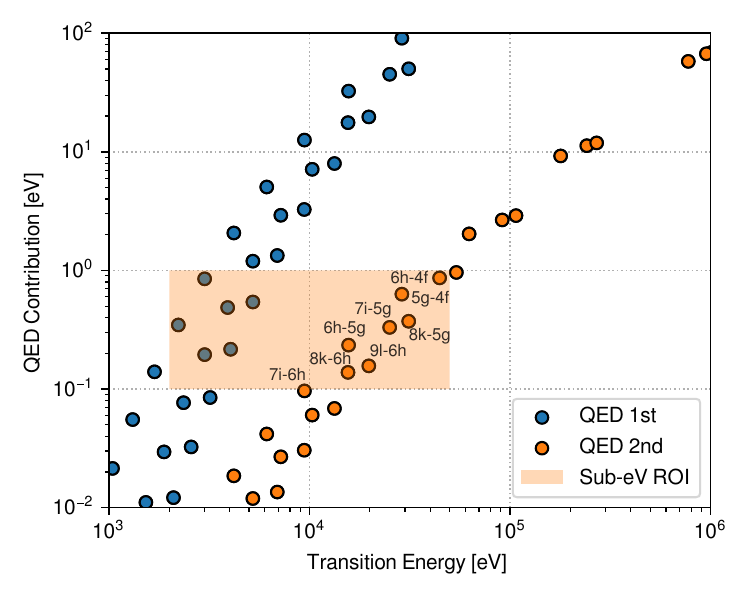}
    \caption{First-order (blue) and second-order (orange) QED contributions as a function of transition energy for K$^{20}$Ne, all in eV. The sub-eV ROI highlights the region sensitive to the SDDs for transition energies in the 2-50 keV range and precisions in the 0.1-1 eV region.}
    \label{fig:BSQED}
\end{figure}

%
% MCDF calculations BSQED plus Isotopic shift 
We performed MCDFGME calculations to compare the experimentally measured transition energies with theoretical predictions. As summarized in Table \ref{tab:results}, the calculated transition energies, $E_{if}^{\text{(calc.)}}$, are presented alongside additional theoretical contributions. Specifically, we report QED contribution $E_{if}^{\text{(QED)}}$ to the transition energies, sum of the first-order $E_{if}^{\text{(QED1)}}$ and second-order $E_{if}^{\text{(QED2)}}$ contributions \cite{indelicato_qed_2019}. Accounting for vacuum polarization diagrams at all-order produces no noticeable effect on the transitions analyzed in this work. We also provide the isotopic shift for each transition, $\Delta E_{if}^\text{(isot.)}$, which was used to constrain the fitting procedure. FNS effects and recoil terms are negligible, of the order of meV, as detailed in the \SM \cite{supp}, where results for different neon isotopes are reported in the Table \ref{tab:extresults}.
% Comparison with experiment
For the most precisely measured line, the 7–6, we find $E_{76}^{\text{(exp.)}} = 9450.23 \pm 0.37\;\mathrm{(stat.)} \pm 1.50\;\mathrm{(syst.)}$ eV, which is in excellent agreement with the calculated value of $E_{76}^{\text{(calc.)}} = 9450.28$ eV where the QED contribution to the transition energy is 12.66 eV.
%, making KNe ideal for BSQED studies.
% Screening effect
We validate these results against the estimation of the energy shift $\Delta E_{if}^{(\text{screen.})}$ due to the screening effect of an electron in the 1s orbital of neon. This contribution, always negative, reflects the reduced effective nuclear charge experienced by the kaon. The precise electron configuration of the atom at the moment of the X-ray transition is not known, since most electrons are expelled during the initial part of the cascade, first from the L shell, then from the K shell \cite{akylas_muonic_1978}. For all studied transitions, electron screening effects were found to be below 1 eV, with the 7-6 line shifted by only $-0.18$ eV. However, at the stage when radiative transitions become dominant ($n < 9$), KNe is expected to be highly ionized, as similarly indicated by cascade calculations performed for KN \cite{koike_electron_2005}. This is further supported by the fact that, for these states, the wavefunctions are well contained within the Ne 1s electron orbital (see FIG. S3 of \SM \cite{supp}), indicating that the atom is fully ionized. These findings establish KNe as a highly effective tool for probing BSQED, owing to its pronounced sensitivity to the magnitude of QED effects.
% Suitability for BSQED Studies 
To further establish the suitability of KNe, and kaonic atoms in general, for BSQED studies, we performed a systematic set of calculations for multiple transitions in KNe up to $n = 15$. Fig. \ref{fig:BSQED} provides an overview of the QED contributions across a range of transitions, emphasizing both the sub-eV precision attainable with SDD detectors and the relevant energy window from 2 to 50 keV for the SIDDHARTA-2 experiment. The survey reveals a variety of QED contributions, with transitions such as 7–6 and 6–5 being particularly prominent. Notably, several additional transitions with $\Delta n = 1,2,3$ where transitions with second-order QED contributions of the order of the sub-eV precision are highlighted.
\subsection{Kaon Mass}
\begin{figure}[t!]
    \centering
    \includegraphics[width=0.99\linewidth]{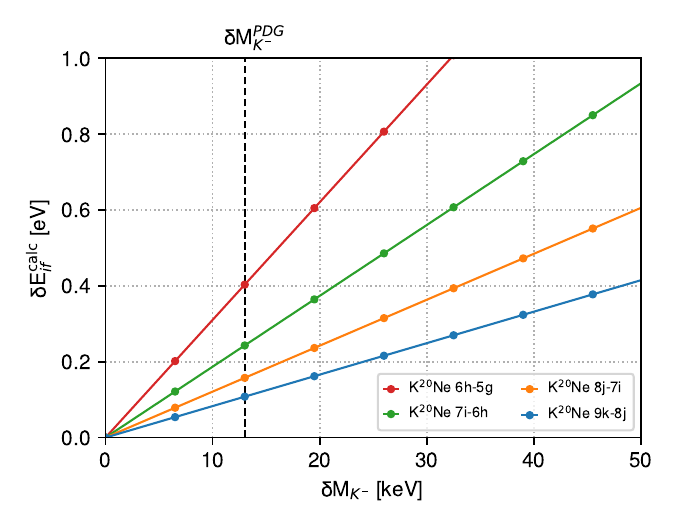}
    \caption{Variation of the transition energies as a function of the kaon mass. The vertical dashed line indicates the current PDG uncertainty of 13 keV on the kaon mass. %The kaon mass value obtained by combining the 76 and 87 transition lines is shown (right) in comparison with other measurements reported by the PDG.
    }
    \label{fig:kmass}
\end{figure}

%
% Sensitivity of calculated lines
We now investigate the impact of the 13 keV PDG kaon mass uncertainty on the size of the BSQED contributions on the KNe lines. We performed several MCDFGME calculations of the transition lines under examination, scaling for the value of the kaon mass. In Fig. \ref{fig:kmass}, we vary the kaon mass by 50 keV with respect to the PDG value and report the variation of the transition energies. Only positive mass variations are shown, as we find the energy-mass relation to be linear at the ppm scale.
%with respect to the values obtained using the PDG one. 
%Moving from higher to lower $n$-transitions, the correction increases, making them more sensitive to the uncertainty on the kaon mass. 
Employing the 13 keV uncertainty from PDG, the theoretical uncertainties in transition energies are listed in the last column of Table \ref{tab:results}. For example, the theoretical uncertainty for the 7–6 transition (0.24 eV) is much smaller than the BSQED contribution (12.66 eV), confirming the suitability of kaonic atoms for BSQED studies despite the current kaon mass uncertainty.
% Kaon Mass extraction
Finally, we show the kaon mass uncertainty achievable with the sub-eV-precision 7–6 and 8–7 transitions of KNe. For each transition, we applied the iterative procedure of Gall \cite{gall_precision_1988}, with full propagation of both statistical and systematic uncertainties. Table \ref{tab:kaonmass} presents the resulting mass values and uncertainties for the 7-6 and 8-7 and their combination. 
%Within the statistical errors, our result already approaches the two most precise values listed in the PDG. This demonstrates that KNe constitute an ideal system for refining the charged‑kaon mass. A dedicated future campaign, with higher statistics and an optimized calibration strategy, could reduce both statistical and systematic uncertainties well below the current PDG value, paving the way for a significantly more precise measurement.
%
\begin{table}[h!]
    \centering
    \caption{Kaon mass ($\MK$) extracted from different transitions, along with statistical ($\delta M_{K^-}^{\text{stat.}}$), systematic ($\delta M_{K^-}^{\text{syst.}}$).} 
    \vskip10pt
    \begin{ruledtabular}
    \begin{tabular}{ccccc}
        % \toprule
        \multirow{2}{*}{\textbf{\boldmath Transition}} & 
        \textbf{\boldmath $\MK$} & 
        \textbf{\boldmath $\delta \MK^{\text{stat.}}$} & 
        \textbf{\boldmath $\delta \MK^{\text{syst.}}$} \\ 
        % \textbf{\boldmath $\delta \MK^{\text{screen.}}$} \\
        % \textbf{\boldmath $\delta \MK$} \\
        & [MeV] & [keV] & [keV] \\%& [keV] \\
        \midrule
        7i-6h & 493.674 & 19 &  78 \\% &  9 \\
        8k-7i & 493.699 & 52 & 121 \\%& 21 \\
        % 9l-8k & 494.208 & 439 & 235 & 44 \\
        \midrule
        7i-6h + 8k-7i & 493.677 & 18 & 66 \\%& 11 \\
        % \bottomrule
    \end{tabular}
    \end{ruledtabular}
    \label{tab:kaonmass}
\end{table}
%
%
%%%%%%%%%%%%%%%%%%%%%%%%%%%%%%%%%%%%%%%%%%%%
\section{Conclusion}
%%%%%%%%%%%%%%%%%%%%%%%%%%%%%%%%%%%%%%%%%%%%
%
%-Purpose Recap
In summary, the comparison of state-of-the-art MCDFGME calculations with recent SIDDHARTA-2 measurements at DA$\Phi$NE establishes kaonic atoms as a robust platform for testing BSQED in intermediate-$Z$ systems. 
%In addition, our results demonstrate that high-yield KNe transitions have potential for being used for the determination of the negatively charged kaon mass.
%-Key Results Summary
From the KNe X-ray spectrum fit, we extracted transition energies for the 7–6, 8–7, and 9–8 lines. The 7–6 and 8–7 transitions achieved sub‑eV statistical precision, with the 7–6 line measured at $9450.23 \pm 0.37\;\mathrm{(stat.)} \pm 1.50\;\mathrm{(syst.)}$ eV, close to the MCDFGME prediction of 9450.28 eV, which includes a QED contribution of 12.66 eV. We also examined two potential systematic effects in the calculations: electron screening and the charged kaon mass uncertainty. For these transitions, the atom is expected to be highly ionized; however, electron screening from a residual 1s electron in neon introduces a shift of only about 1\% of the QED contribution. Moreover, the current PDG uncertainty of 13 keV on the kaon mass induces only a few percent impact on transition energies. Together, these findings highlight that theoretical uncertainties remain at the percent level relative to the dominant QED contribution, making kaonic atoms ideal systems for precision BSQED tests in the intermediate‑Z regime. 
In the end, we have shown that by combining the sub-eV-precision 7–6 and 8–7 transitions, it is possible to achieve a statistical uncertainty on the kaon mass below 20 keV, already approaching the precision of the two most accurate, but mutually inconsistent, measurements reported by the PDG. We note that the present experimental setup was optimized for the deuterium measurement; a dedicated campaign focused on kaonic neon could bring the total uncertainty on the kaon mass below 10 keV. This would be achievable by reducing systematic uncertainties through dedicated calibration of the KNe lines at 6 keV and 9 keV. Additionally, doubling the statistics and optimizing the kaon stopping efficiency specifically for a neon gas target would further lower the statistical uncertainty, potentially reaching the sub-10 keV level.
%-Future Work / Outlook
Our results establish KNe as a benchmark for future BSQED tests and for improved determinations of fundamental hadronic properties. Further advances in detector technology and statistics, development of more refined cascade models, and the extension of this methodology to other intermediate-$Z$ kaonic atoms promise even more stringent tests of QED in strong fields. Such progress will enable improved determinations of fundamental particle properties, including the charged kaon mass, and may shed light on discrepancies in precision spectroscopy, hadronic interactions, and the interplay between QED and the strong force in exotic atoms.
%
%%%%%%%%%%%%%%%%%%%%%%%%%%%%%%%%%%%%%%%%%%%%
\section{Acknowledgments}
%%%%%%%%%%%%%%%%%%%%%%%%%%%%%%%%%%%%%%%%%%%%
%
We thank C. Capoccia from INFN-LNF and H. Schneider, L. Stohwasser, and D. Pristauz-Telsnigg from Stefan Meyer-Institut for their fundamental contribution in designing and building the SIDDHARTA-2 setup. We also thank INFN-LNF and the DA$\Phi$NE staff for the excellent working conditions and their ongoing support. Special thanks to Catia Milardi for her continued support and contribution during the data taking. We gratefully acknowledge Polish high-performance computing infrastructure PLGrid (HPC Center: ACK Cyfronet AGH) for providing computer facilities and support within computational grant no. PLG/2025/018524. Part of this work was supported by the INFN (KAONNIS project); the Austrian Science Fund (FWF): [P24756-N20 and P33037-N]; the Croatian Science Foundation under the project IP-2022-10-3878; the EU STRONG-2020 project (Grant Agreement No. 824093); the EU Horizon 2020 project under the MSCA (Grant Agreement 754496); the Japan Society for the Promotion of Science JSPS KAKENHI Grant No. JP18H05402, JP22H04917; the Polish Ministry of Science and Higher Education grant No. 7150/E-338/M/2018 and the Polish National Agency for Academic Exchange (grant no PPN/BIT/2021/1/00037); the EU Horizon 2020 research and innovation programme under project OPSVIO (Grant Agreement No. 101038099).
%
%%%%%%%%%%%%%%%%%%%%%%%%%%%%%%%%%%%%%%%%%%%%
\section{Data Availability}
%%%%%%%%%%%%%%%%%%%%%%%%%%%%%%%%%%%%%%%%%%%%
%
Data and input files for the MCDFGME code are available upon reasonable request.
% \nocite{*}
% \section*{References}
\bibliographystyle{apsrev4-2}
\bibliography{Kaonic_Neon}% Produces the bibliography via BibTeX.
% \printbibliography[heading=none]
% \vskip100pt
\clearpage
\onecolumngrid
\begin{center}
    \textbf{\large Supplemental Material:\\
    Precision Test of Bound-State QED at Intermediate-Z with Kaonic Neon}
\end{center}
\vspace{0.5cm}
%
%%%%%%%%%%%%%%%%%%%%%%%%%%%%%%%%%%%%%%%%%%%%
\section*{S1. Extended Table of MCDFGME Transition Energies}
%%%%%%%%%%%%%%%%%%%%%%%%%%%%%%%%%%%%%%%%%%%%
%
\renewcommand{\thefigure}{S\arabic{figure}}
\setcounter{figure}{0}
\renewcommand{\thetable}{S\arabic{table}}
\setcounter{table}{0}
\renewcommand{\theequation}{S\arabic{equation}}
\setcounter{equation}{0}
\begin{table*}[h!]
    \centering
    \caption{MCDFGME Calculations for K\textsuperscript{20}Ne and K\textsuperscript{22}Ne for different lines, with QED contributions, FNS and Recoil effects, electron screening and uncertainty due to PDG kaon mass. All energies are given in eV.}
    \vskip10pt
    \begin{ruledtabular}
    \begin{tabular}{ccccccccccc}
        \textbf{Isotope} & 
        \textbf{Transition} &
        \textbf{\boldmath $E_{if}^{\text{(calc.)}}$} &
        \textbf{\boldmath $E_{if}^{\text{(QED)}}$} & 
        \textbf{\boldmath $E_{if}^{\text{(QED1)}}$} & 
        \textbf{\boldmath $E_{if}^{\text{(QED2)}}$} &
        \textbf{FNS} &
        \textbf{Recoil} &
        \textbf{\boldmath $\Delta E_{if}^{\text{(screen.)}}$} &
        \textbf{\boldmath $\Delta E_{if}^{\text{(PDG)}}$} \\
        \midrule
        \multirow{4}{*}{K\textsuperscript{20}Ne} 
        & 9l-8k  & 4201.45  & 2.09 & 2.07  & 0.02 & 0.0003 & 0.0042 & -0.38 & 0.11 \\
        & 8k-7i  & 6130.31  & 5.09 & 5.05  & 0.04 & 0.0008 & 0.0079 & -0.27 & 0.16 \\
        & 7i-6h  & 9450.28  & 12.66 & 12.56 & 0.10 & 0.0026 & 0.0162 & -0.18 & 0.24 \\
        & 6h-5g  & 15685.39 & 32.74 & 32.51 & 0.23 & 0.0104 & 0.0375 & -0.11 & 0.40 \\
        \midrule
        \multirow{4}{*}{K\textsuperscript{22}Ne} 
        & 9l-8k  & 4211.35  & 2.10 & 2.08  & 0.02 & 0.0003 & 0.0039 & -0.38 & 0.11 \\
        & 8k-7i  & 6144.76  & 5.13 & 5.09  & 0.04 & 0.0008 & 0.0073 & -0.27 & 0.16 \\
        & 7i-6h  & 9472.57  & 12.73 & 12.63 & 0.10 & 0.0026 & 0.0149 & -0.18 & 0.24 \\
        & 6h-5g  & 15722.40 & 32.91 & 32.67 & 0.24 & 0.0101 & 0.0345 & -0.11 & 0.41 \\
    \end{tabular}
    \end{ruledtabular}
    \label{tab:extresults}
\end{table*}
%
%
%%%%%%%%%%%%%%%%%%%%%%%%%%%%%%%%%%%%%%%%%%%%
\section*{S2. Isotopic Shift as a function of Kaon Mass}
%%%%%%%%%%%%%%%%%%%%%%%%%%%%%%%%%%%%%%%%%%%%
%
%
\begin{figure*}[h!]
    \centering
    \includegraphics[width=0.99\linewidth]{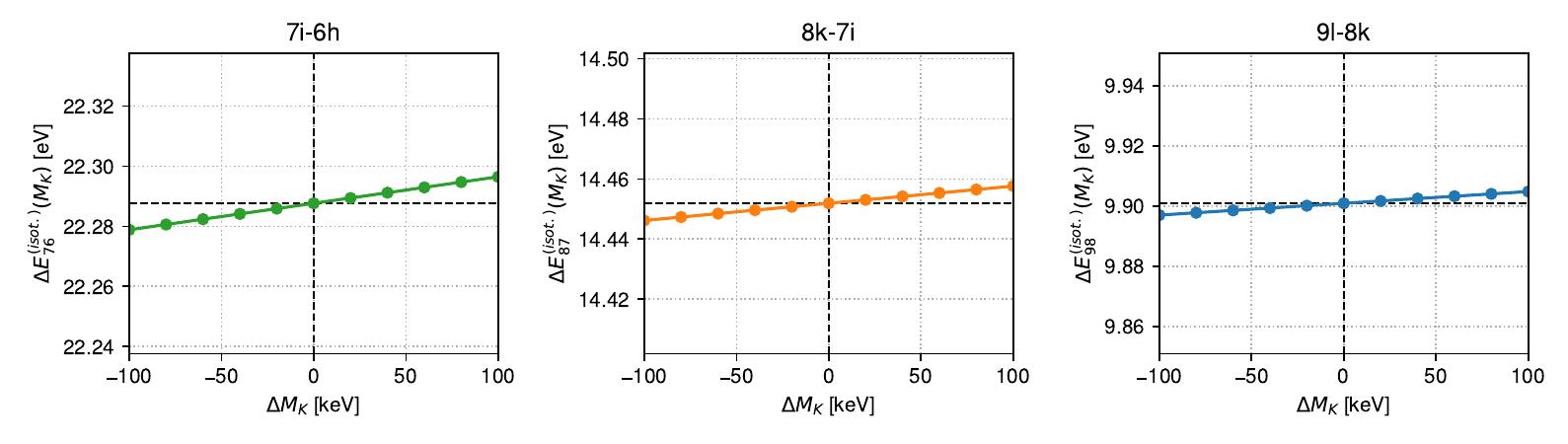}
    \caption{Sensitivity of isotopic energy shifts to the kaon mass for the $7i$–$6h$, $8k$–$7i$, and $9l$–$8k$ transitions in K\textsuperscript{20}Ne. Each panel shows the variation of $\Delta E^{(\mathrm{isot.})}_{n,n-1}(M_K)$ as a function of $\Delta M_K$. The maximum variation over the $\pm 100$keV range is 9 meV for $7i$–$6h$, 6 meV for $8k$–$7i$, and 4 meV for $9l$–$8k$.}
    \label{fig:isotopeshift}
\end{figure*}
%
%
%%%%%%%%%%%%%%%%%%%%%%%%%%%%%%%%%%%%%%%%%%%%
\section*{S3. Radiative and Auger Rates}
%%%%%%%%%%%%%%%%%%%%%%%%%%%%%%%%%%%%%%%%%%%%
%
The formulas used to compute the radiative and Auger rates for KNe are taken from \cite{burbidge_mesonic_1953}:
\begin{align}
    &\Gamma_{n,\,l\rightarrow n',\,l\pm 1}^{R} = \frac{4\mu Z^4}{3} \alpha^3 \left| R_{n',\,l\pm 1}^{n,\,l} \right|^{2} (\Delta E_{if})^3 \\
    &\Gamma_{n,\,l\rightarrow n',\,l\pm 1}^{A} = \frac{16}{3} \left( \frac{Z_e}{Z} \right)^{2} \frac{\pi}{\mu^2} \frac{l}{2l+1} \left| R_{n',\,l\pm 1}^{n,\,l} \right|^{2} \frac{y^2}{1 + y^2} \frac{\exp \left[ y (4\tan^{-1} y - \pi) \right]}{\sinh(\pi y)}.
\end{align}
The Auger rate formula applies for an electron in the $1s$ orbital, and $Z_e=Z-1$. $\Delta E_{if}$ is the transition energy for the hydrogenic case and $R_{n',\,l\pm 1}^{n,\,l}$ is the dipole radial integral. $\mu$ denotes the reduced mass of the K$^{20}$Ne system. Here, $y = \sqrt{k/2E}$ is the momentum of the ejected electron.
Fig. S2 compares the radiative transition rates computed using the scaled hydrogenic formula [Eq. (3)] and those obtained from MCDFGME calculations, for transitions between circular states in KNe. For most high-$n$ transitions ($n>5$), the agreement is better than $1\%$, with the largest deviation observed for the $2p \rightarrow 1s$ transition. Contributions from higher multipoles are negligible for these transitions.
\begin{figure*}[h!]
    \centering
    \includegraphics[width=0.7\linewidth]{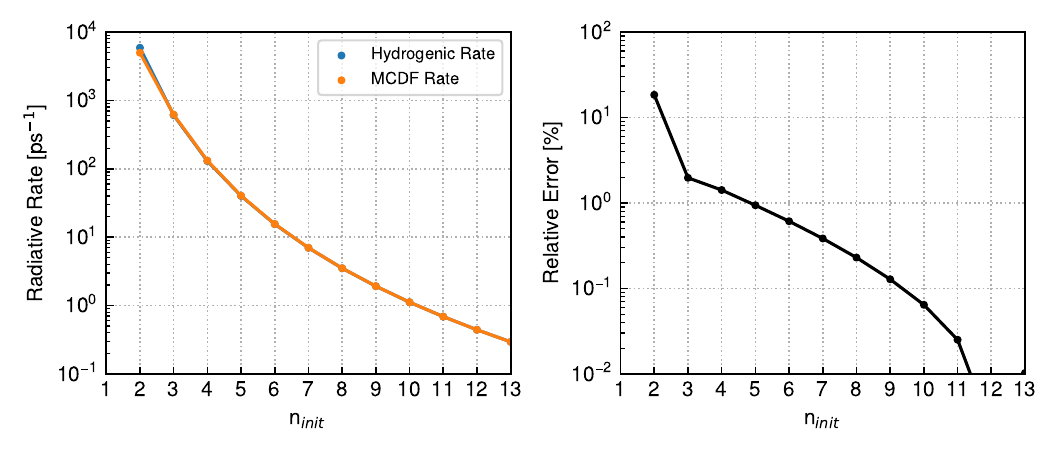}
    \caption{Comparison of radiative transition rates between circular states in KNe calculated with the scaled hydrogenic formula \cite{burbidge_mesonic_1953} (blue) and from MCDFGME wavefunctions (orange). The agreement is within $2\%$ for most high-$n$ transitions; significant deviations occur mainly for the $2p\rightarrow 1s$ line. Multipole contributions are found to be negligible in this regime.}
    \label{fig:radtest}
\end{figure*}
%
%
%%%%%%%%%%%%%%%%%%%%%%%%%%%%%%%%%%%%%%%%%%%%
\section*{S4. Kaon and Electron Wavefunction}
%%%%%%%%%%%%%%%%%%%%%%%%%%%%%%%%%%%%%%%%%%%%
%
%
\begin{figure*}[h!]
    \centering
    \includegraphics[width=0.7\linewidth]{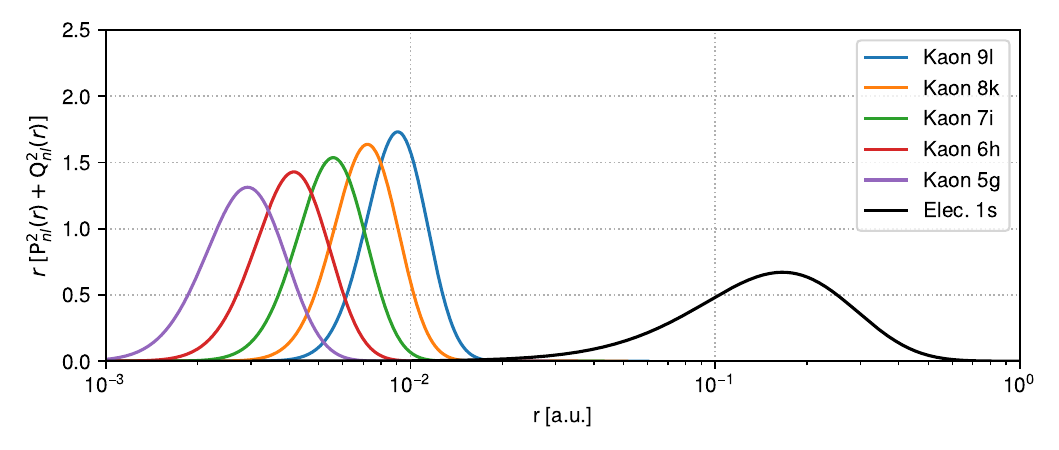}
    \caption{Radial probability densities for selected kaonic neon states (9l, 8k, 7i, 6h, 5g) compared with the electron 1s wavefunction of neon. The kaonic wavefunctions are entirely contained within the electron 1s shell, demonstrating that the kaon orbits deep inside the electronic cloud. In this regime, the atom is fully ionized and electron screening effects are negligible.}
    \label{fig:wf}
\end{figure*}
%
%%%%%%%%%%%%%%%%%%%%%%%%%%%%%%%%%%%%%%%%%%%%
\section*{S5. Iterative Determination of the Kaon Mass}
%%%%%%%%%%%%%%%%%%%%%%%%%%%%%%%%%%%%%%%%%%%%
%
%
\begin{figure*}[h!]
    \centering
    \includegraphics[width=0.7\linewidth]{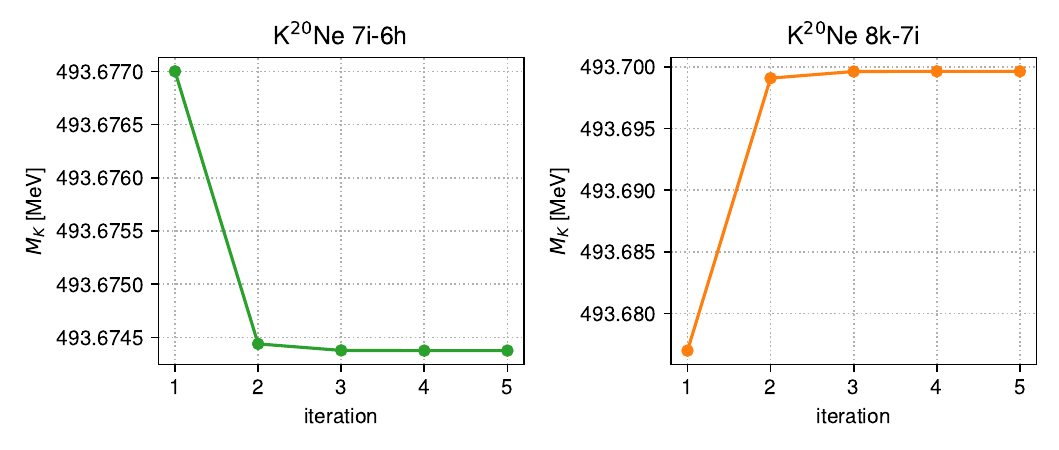}
    \caption{Evolution of the extracted kaon mass, $M_K$, in MeV, as a function of iteration for the 7i-6h and 8k-7i transitions in K$^{20}$Ne. The threshold for terminating the mass iteration was set to 0.5 keV.}
    \label{fig:kaon_mass_iterations}
\end{figure*}
%
%
%%%%%%%%%%%%%%%%%%%%%%%%%%%%%%%%%%%%%%%%%%%%
% \section*{S6. Degrader Curves}
%%%%%%%%%%%%%%%%%%%%%%%%%%%%%%%%%%%%%%%%%%%%
%
%
% \begin{figure*}[h!]
%     \centering
%     \includegraphics[width=0.7\linewidth]{fs4_Degrader_Curve.pdf}
%     \caption{Yields for the 76 and 87 transitions in KNe as a function of the degraded thickness.}
%     \label{fig:kaon_mass_iterations}
% \end{figure*}
%
%
\end{document}